# Nanoscale control of perpendicular magnetic anisotropy, interfacial Dzyaloshinskii-Moriya interaction and skyrmions in the inversion-symmetry-broken Ru/Co/W/Ru films

Kolesnikov A.G.[1], Ognev A.V.[1], Stebliy M.E.[1], Chebotkevich L.A.[1], Gerasimenko A.V.[2], Sadovnikov A.V.[3,4],

Nikitov S.A.[3,4], Samardak A.S.[1,5]*

[1]School of Natural Sciences, Far Eastern Federal University, Vladivostok, Russia
[2]Institute of Chemistry, Far East Branch, Russian Academy of Sciences, Vladivostok, Russia
[3]Laboratory "Metamaterials", Saratov State University, Saratov, Russia
[4]Kotel'nikov Institute of Radioengineering and Electronics, Russian Academy of Sciences, Moscow, Russia
[5]Center for Spin-Orbitronic Materials, Korea University, Seoul, Republic of Korea
e-mail address: samardak.as@dvfu.ru

An enhancement of the spin-orbit effects arising on an interface between a ferromagnet (*FM*) and a heavy metal (*HM*) is possible through the strong breaking of the structural inversion symmetry in the layered films. Here we show that the introduction of an ultrathin W interlayer between Co and Ru in Ru/Co/Ru films enables to preserve perpendicular magnetic anisotropy (PMA) and simultaneously induce a large interfacial Dzyaloshinskii-Moriya interaction (iDMI). We find that the Ru/Co/W/Ru films have PMA up to 0.35 nm of the nominal thickness of W ($t_W$). The study of the spin-wave propagation in the Damon-Eshbach geometry by Brillouin light scattering (BLS) spectroscopy reveals the drastic increase of the iDMI value with the rising $t_W$. The maximum iDMI of -3.1 erg/cm$^2$ is observed for $t_W$=0.24 nm, which is 10 times larger than the latter for the quasi-symmetrical Ru/Co/Ru films. The ability to simultaneous control the strength of PMA and iDMI in symmetrical *HM/FM/HM* trilayer systems through the interface engineered inversion asymmetry at the nanoscale excites new fundamental and practical interest to the chiral ferromagnets, which are a potential host for magnetic skyrmions.

## I. INTRODUCTION

Recently, in perpendicularly magnetized heterostructures with *FM/HM* interfaces a number of novel spin-orbit effects, which are prospective for spintronic applications, has been observed [1-3]. Strong spin-orbit coupling between *FM* and *HM* atoms induces spin-Hall effect-driven spin-orbit torque (SOT) [4, 5], anisotropic chiral damping [6], spin-wave canting [7], indirect exchange coupling and Dzyaloshinskii-Moriya interaction (DMI) [8, 9]. These spin orbital effects are promising for non-volatile memory and artificial intelligence applications [10], like SOT-MRAM [11, 12], racetrack memory [13, 14], and neuromorphic computing [15-17]. For instance, SOT enables to switch the out-of-plane magnetization of a magnetic layer by current passing through a heavy metal. Noteworthy, the required current density is significantly smaller if to compare with the spin transfer torque (STT) driven magnetization switching [11]. Anisotropic chiral damping defines the velocity of domain walls in dependence on the propagation direction. The interfacial DMI, an anti-symmetric exchange arising on *FM/HM* interfaces, stabilizes the homochiral Neel domain walls and topologically protected spin textures, like skyrmions and merons [3, 18].

Since we discuss the spin-orbit effects at the interfaces, than their action strongly exposes in ultrathin ferromagnetic layers [19-21]. In case of a trilayer system, like $HM_1/FM/HM_2$, DMI arising on the top and bottom interfaces can lead to its cancelation or enhancement in dependence on the interaction sign. The complete compensation of DMI has to be observed in films with the ideally symmetrical interfaces. However, quasi-symmetrical trilayers, like Pt/Co/Pt, possess a non-zero DMI due to the different quality of the top and bottom interfaces [22-24]. Recently, the impact of the breaking of the structural inversion symmetry on the interfacial DMI has been studied in a wide range of systems: Pt/Co/AlO$_x$ [25], Pt/CoFeB/AlO$_x$ [26], Ir/Co/AlO$_x$ [27], Pt/Co/MgO [28], Pt/CoFeB/MgO [29], Ta/CoFeB/TaO$_x$ [30], Au/Co/W [31], Ir/Co/Pt [27, 32], Ir/Fe/Co/Pt [33], Pt/Co/Ni/Ta [4]. Nowadays, the most promising interface engineering approach is based on the incorporation of an ultrathin metallic or non-metallic interlayer, calling a dusting layer, into an interface between a ferromagnet layer and a heavy metal or an oxide capping: Pt/Co/*Ir*/Pt [20], Pt/Co/*Cu*/AlO$_x$ [34], Pt/Co/(*W,Ta,Pd*)/Pt

[35], Pt/*Cu*/Co/Pt [36], MgO/Co/*Cu*/Pt/Ta [37], Pt/Co/*C*/Ta [38].

In our previous paper [39] we demonstrated magnetic properties of quasi-symmetrical Ru/Co/Ru trilayers and unveiled the layer composition in order to get strong perpendicular magnetic anisotropy (PMA), which value can be tuned by the thickness of the Ru buffer. However, our preliminary study has showed that this structure has weak spin-Hall effect ($\Theta_{SH}{\approx}0.005$) and small iDMI ($D_{eff}$=-0.27 erg/cm$^2$). To enhance the spin orbit effects, we propose to use the ultrathin interlayer of a heavy metal with the opposite to Ru signs of $\Theta_{SH}$ and iDMI. This interlayer has to be introduced into the top Co/Ru interface, because the bottom Ru/Co interface induces the out-of-plane magneto-elastic anisotropy and has to be preserved from any structural modification. The most suitable material for our task is tungsten (W), because it has the negative spin Hall angle ($\Theta_{SH}$=-0.14) [40] and positive iDMI revealed by our measurements on W/CoFeB/MgO trilayers. Our preliminary results showed that the complete substitution of the Ru capping by W led to vanishing of PMA and decreasing of saturation magnetization. To keep PMA and enforce the structural asymmetry, we introduce into the Co/Ru interface a dusting W interlayer with nominal thickness $t_w$ ranging from 0 to 0.4 nm.

In this paper, we study the influence of the Co and W thicknesses on the magnetic properties, magnetization reversal, iDMI and skyrmion stabilization in the inversion-symmetry-broken Ru/Co/W/Ru films.

## II. EXPERIMENTAL

Our polycrystalline Ru/Co/W/Ru films were prepared by magnetron sputtering on the SiO$_2$ substrates at room temperature. The base pressure in the chamber was 10$^{-8}$ Torr. The working pressure of $Ar^+$ was 10$^{-4}$ Torr. In order to precise control the thickness of layers, we used low sputtering rates: $V(Ru)$=0.011 nm/s, $V(Co)$=0.018 nm/s, $V(W)$=0.02 nm/s. The Co thickness ($t_{Co}$) was varied from 0.7 to 1.5 nm. The thickness of the buffer and capping Ru layers ($t_{Ru}$) was 10 and 2 nm, correspondingly. The W thickness ($t_W$) was taken in the range from 0 to 0.4 nm. The structural and magnetic properties of the quasi-symmetrical Ru/Co/Ru were systematically studied in [39].

Magnetic properties of films were investigated with vibrating sample magnetometer (7410 VSM, LakeShore). Kerr microscopy (Evico Magnetics) was used to study domain structure and its field driven dynamics. Brillouin light scattering (BLS) spectroscopy measurements were conducted to probe the spin-wave frequency non-reciprocity in the Damon-Eshbach geometry in order to define the iDMI sign and value. The crystal structure and interface quality was studied by X-ray diffraction (XRD) and X-ray reflectivity (XRR) measurement techniques (SmartLab, RIGAKU) at CuKα radiation wavelength (1.54Å). The fitting of XRR spectra were performed with GenX software [41]. Micromagnetic simulations of the domain structure were executed using MuMax$^3$ software package [42].

## III. PERPENDICULAR MAGNETIC ANISOTROPY

The main purpose of our study is to significantly enhance the structural inversion asymmetry and, consequently, iDMI, simultaneously keeping PMA in the films. Our previous results on Ru/Co/Ru trilayers show that PMA arises due to the elastic stress in Co caused by the lattice mismatch (7.8%) between Co and Ru [39]. We assume that the introduction of another heavy metal on the top of the Co surface will change the intrinsic strains in Co and, consequently, modify the PMA value. Firstly, we defined thicknesses of Co in Ru/Co/W films with the fixed nominal thickness of W $t_W$=0.23 nm, at which PMA realizes.

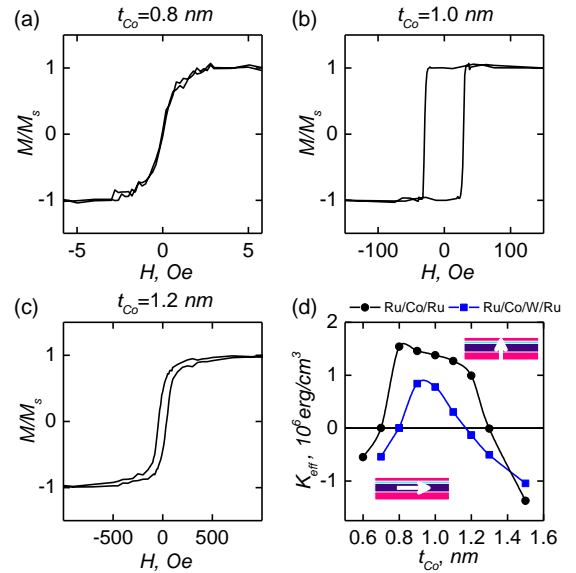

**FIG.1.** Magnetic hysteresis loops measured by VSM in the out-of-plane geometry for Ru(10)/Co($t_{Co}$)/W(0.23)/Ru(2) films with different Co thicknesses: (a) $t_{Co}$=0.8 nm, (b) 1.0 nm, (c) 1.2 nm. (d) Dependences of the effective magnetic anisotropy energy $K_{eff}$ on the Co layer thickness for Ru/Ru (black curve) and Ru/Co/W/Ru (blue curve) films.

Figure 1(a-c) demonstrates the out-of-plane hysteresis loops for Ru(10)/Co($t_{Co}$)/W(0.23)/Ru(2) films with $t_{Co}$=0.8, 1.0 and 1.2 nm, correspondingly. At $t_{Co}$=1.0 nm the loop has rectangular shape with maximum value of the reduced remanent magnetization $M_r/M_s$=1 (Fig.1(b)) pointing out that the easy axis of magnetization (EAM) is perpendicular to the sample plane. As seen in Fig.1(d), PMA is observed for $t_{Co}$ ranging from 0.8 to 1.17 nm, where the effective magnetic anisotropy energy ($K_{eff}$) is positive. Outside this range ($t_{Co}{\leq}0.8$ nm and $t_{Co}{\geq}1.17$ nm), $K_{eff}{\leq}0$ and EAM lies in the sample plane. For comparison, we show in Fig.1(d) the $K_{eff}$=$f(t_{Co})$ curve for Ru/Co/Ru films, where one can see that the introduction of W leads to decrease of the PMA existence range and the effective anisotropy energy.

# IV.   STRUCTURAL MODIFICATION OF THE TOP INTERFACE

To unveil the microscopic origin of the change of PMA in Ru/Co/W/Ru films, we studied the influence of the W dusting interlayer on the structural quality of the top Co/Ru interface by XRR.

Due to the atom interdiffusion during sputtering, magnetically dead layers can appear on the bottom and top interfaces [43, 44]. The thickness of the dead layers ($\Lambda = \Lambda_{top} + \Lambda_{bottom}$) depends on the contacting materials [45]. We defined $\Lambda$ as a cross point of a $M_S \cdot t_{Co} = f(t_{Co})$ line with the $x$ axis, Fig.2(a). For the quasi-symmetrical Ru/Co and Co/Ru interfaces, the total thickness of both dead layers ($\Lambda_{Co/Ru} + \Lambda_{Ru/Co}$) was 0.4 nm. This value was proved by the XRR study, Fig.2(c). This method allows to deduce the thickness and density of each layer, roughness of interfaces and interdiffusion depth in thin multilayer films [46]. The analysis of the experimentally measured XRR spectra was conducted in GenX free software package [41]. From the fitting of the experimental data we deduced the values of $\Lambda$ and the root mean square roughness ($\sigma$) of the top and bottom interfaces: $\Lambda_{Ru/Co} = 0.198$ nm and $\sigma_{Ru/Co} = 0.32$ nm, $\Lambda_{Co/Ru} = 0.201$ nm and $\sigma_{Co/Ru} = 0.6$ nm - for Ru(10)/Co(1)/Ru(2); $\Lambda_{Ru/Co} = 0.199$ nm and $\sigma_{Ru/Co} = 0.31$ nm; $\Lambda_{Co/W} = 0.43$ nm and $\sigma_{Co/W} = 0.28$ nm – for Ru(10)/Co(1)/W(0.23)/Ru(2).

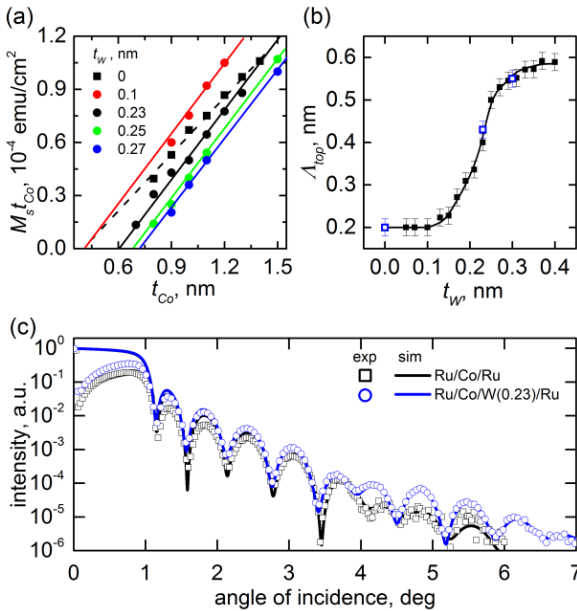

**FIG.2.** (a) Linear dependences of $M_S \cdot t_{Co} = f(t_{Co})$ for Ru(10)/Co($t_{Co}$)/W($t_W$)/Ru(2) films with variation of $t_W$. (b) Thickness of the magnetically dead layer on the top interface ($\Lambda_{top}$) as a function of $t_W$ (black and blue squares correspond to the data deduced from (a) and from XRR study). (c) XRR spectra for Ru(10)/Co(1)/Ru(2) (black) and Ru(10)/Co(1)/W(0.23)/Ru(2) (blue), where dots and lines match experimental and simulated data, respectively.

Thus, the W interlayer promotes the increase of $\Lambda$ and the decrease of $\sigma$ on the top interface of Co. At this the $\Lambda$ value depends on $t_W$: the increase of $t_W$ from 0.1 to 0.35 nm leads to the growth of $\Lambda_{Co/W}$ from 0.2 to 0.58 nm, Fig.2(b). We suggest that the diffusion of W atoms into Co layer possibly favors to formation of $Co_3W$ and $Co_7W_6$ phases, which have the $hcp$ lattice. As a result, the lattice mismatch on the top interface decreases down to 0.53 and 1.2 % for Co/$Co_3W$ and Co/$Co_7W_6$ [47-49], correspondingly, meanwhile the lattice mismatch on the bottom Ru/Co interface remains 7.8%.

Since the introduction of the W interlayer into the top interface increases the magnetically dead layer, the effective thickness of the Co layer decreases. Structural changes on the Co/W interface diminish the elastic strains in Co resulting in the decrease of PMA.

# V.   EFFECT OF THE W INTERLAYER ON MAGNETIC PROPERTIES

The structural modification of the top interface have to change the magnetic properties of the films. Since, we are interested in systems with PMA, the thickness of the ferromagnetic layer $t_{Co} = 1$ nm was chosen accordingly to the $K_{eff} = f(t_{Co})$ plot (Fig.1(d)), blue curve in the middle of the range, where $K_{eff} > 0$ and PMA has the maximum value. At this fixed thickness of Co, we investigated the effect of $t_W$ ranging from 0 to 0.4 nm on the magnetic behavior. The VSM study shows that the out-of-plane hysteresis loops keep their rectangular shape up to $t_W = 0.35$ nm (Fig. 3a). The Co/W interface roughness decreases with the increasing $t_W$, which promotes the fall of the coercive field ($H_c$), Fig.3(b). We observed the same linear decrease of $H_c$ for the films with $t_{Co} = 0.9$ and 1.1 nm. To define the $K_{eff}$ values (Fig.3(d)), we measured the in-plane hysteresis loops, where the field was applied along the hard axis of magnetization, Fig.3(c). The introduction of the ultrathin W layer with $t_W = 0.05$ nm leads to the abrupt fall of $K_{eff}$ from 0.7 down to 0.55 ×$10^6$ erg/cm$^3$ ($\Delta K_{eff} = 0.15 \times 10^6$ erg/cm$^3$), which also affects on the $H_c$ value. However, at the further increase of $t_W$ up to 0.21 nm, the $K_{eff}$ value remains constant.

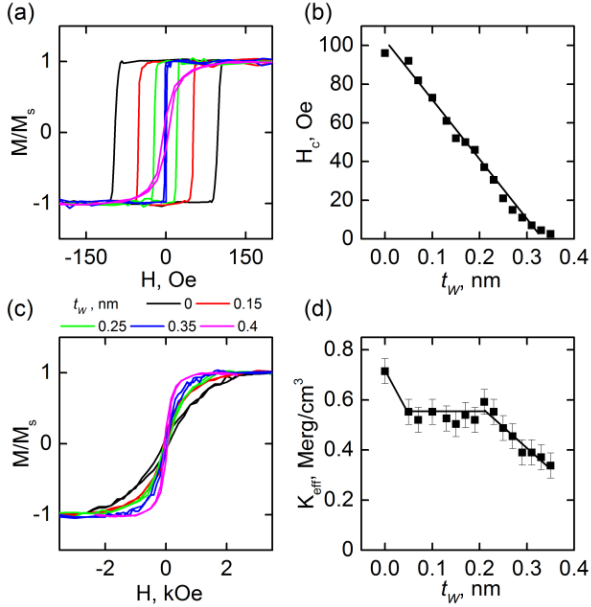

**FIG.3.** Magnetic hysteresis loops recorded in the field applied in the out-of-plane (a) and in-plane (c) geometries for Ru(10)/Co(1)/W($t_W$)/Ru(2). (b) Dependence of $H_c$ on $t_w$. (d) The effective anisotropy energy $K_{eff}$ as a function of $t_w$.

In polycrystalline films, the effective energy of magnetic anisotropy consists of the volume contribution $K_v=-2\pi M_s^2 + K_{me}$, caused by the shape ($2\pi M_s^2$) and magneto-elastic ($K_{me}$) anisotropies, and of the surface contribution from the bottom and top interfaces, $K_{S1}$ and $K_{S2}$, correspondingly:

$$K_{eff}=-2\pi M_s^2 + K_{me}+(K_{S1}+K_{S2})/t_{Co}  \quad (1)$$

The modification of the Co/Ru interface by the W dusting changes the surface contribution $K_{S2}$ into $K_{eff}$. In our work [39] the energy of $K_{S2}=0.9\times10^{-3}$ erg/cm$^2$ for the top Co/Ru interface without W was determined experimentally and calculated theoretically. The ratio of the surface contribution of the bottom interface to the Co thickness taking into account the magnetically dead layer was $K_{S2}/(t_{Co}-(\Lambda_{Ru/Co}+\Lambda_{Co/Ru}))=0.15\times10^6$ erg/cm$^3$. This value matches the change of $\Delta K_{eff}$. The decrease of $K_{eff}$ from 0.55 down to 0.34 $\times10^6$ erg/cm$^3$ for $t_w$ ranging from 0.21 to 0.35 nm is related to the reduction of the elastic strains in the ferromagnetic layer and to the descent of the magneto-elastic anisotropy energy $K_{me}$, Fig.3.(d).

The modified magnetic properties of Ru/Co/W($t_W$)/Ru films are reflected on their magnetic domain structure. Magnetization reversal process in a magnetic field applied perpendicular to the sample plane was studied by Kerr microscopy. In the quasi-symmetrical Ru/Co/Ru films magnetic domains nucleate in the field a little less than $H_c$=-95 Oe, Fig.4(a). With the increasing field, domains grow due to the domain wall displacement with the following domains collation resulting in the complete magnetization switching.

An introduction of the 0.35 nm-thick W interlayer decreases the coercive field ($H_c$=2.5 Oe) and changes the magnetization reversal mode, Fig.4(b). The small domains nucleate in $H$=-1.5 Oe. The rising field causes in the increasing number of domains and the dendrite-like structure formation. However, the approaching domains, having homochiral 180° Neel domain walls, do not collate, but form 360° domain walls. The formation of the homochiral domain walls in films with a single ferromagnetic layer is a fingerprint of the presence of the Dzyaloshinskii-Moriya interaction [21].

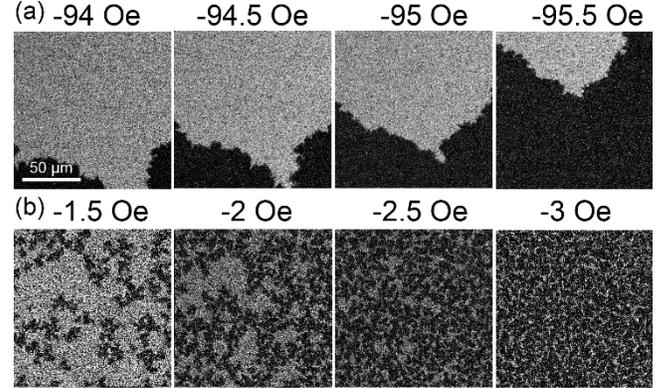

**FIG.4.** Magnetic domain structure of (a) Ru(10)/Co(1)/Ru(2) and (b) Ru(10)/Co(1)/W(0.35)/Ru(2) films, captured by the polar Kerr effect in the out-of-plane geometry.

# VI. INTERFACIAL DZYALOSHINSKII-MORIYA INTERACTION

To extract the average iDMI values measured for large film areas, we used the Brillouin light scattering (BLS) spectroscopy of non-reciprocal propagation of spin waves in the Damon-Eshbach geometry, when a sample is in-plane magnetized and the wave vector is perpendicular to the magnetization [50]. The shift between the Stokes and anti-Stokes frequencies of spin waves of a given wavelength is directly proportional to the iDMI value [26]:

$$\Delta f = f_S - f_{AS} = \frac{2\gamma}{\pi M_s} D_{eff} k_x = \frac{2\gamma D_S}{\pi M_s t_{FM}} k_x,  \quad (2)$$

where $f_S$ and $f_{AS}$ are the Stokes and anti-Stokes frequencies, correspondingly, $\gamma$ is the gyromagnetic ratio, $M_s$ is the saturation magnetization, $k_x$ is the spin wave vector, $t_{FM}$ is the effective ferromagnetic layer thickness considering the total thickness of magnetically dead layers of the bottom and top interfaces ($t_{FM}=t_{Co}-(\Lambda_{Ru/Co}+\Lambda_{Co/W})$), $D_{eff}$ is the effective (thickness-averaged) energy density of the Dzyaloshinskii-Moriya interaction given in [erg/cm$^2$], $D_S=D_{eff}t_{FM}$ is the thickness-independent surface DMI constant taking into account the contributions from the bottom and top interfaces and it is given in [erg/cm].

Figure 5(a) demonstrates the typical BLS spectra for Ru(10)/Co(1)/W(0.25)/Ru(2) film recorder for $k_x$=11 µm$^{-1}$ ($\theta\approx28°$) in the positive and negative fields $H=\pm5$ kOe. If a film has DMI, the spin wave frequencies will depend on the spin waves propagation direction at the fixed wave vector of the incident light as $k_x = \dfrac{4\pi}{\lambda}\sin\theta$, where $\lambda$ is the wavelength of the light (532 nm). The effective value of DMI can be defined from (2) as:

$$D_{eff} = \frac{\pi l M_s \Delta f}{2\gamma k_x},\qquad(3)$$

where $\gamma = \dfrac{g\mu_B}{\hbar} = g \times 87.94\ GHz/T$, $g$ is the Landé factor (for ultrathin Co films $g$=2.14 [51], $\gamma$=188 GHz/T). We found for Ru(10)/Co(1)/W(0.25)/Ru(2) film with $M_s$=1300 emu/cm$^3$ that $D_{eff}$=-2.71 erg/cm$^2$ and $D_s$=-0.84 $10^{-7}$ erg/cm, which values are comparable or even higher than that for the early measured films: Pt/Co/AlO$_x$ [25, 27], Pt/Co/GdO$_x$ [52], Pt/CoFe/MgO [53] and Pt/Co/Fe/Ir [33]. The negative sign of iDMI supposes that the left-handed cycloidal spin structures are most feasible [54, 55]. The interfacial nature of the define DMI was confirmed by the observed linear dependence of $\Delta f$ on the sample tilt angle $\theta$ (or spin wave vector $k_x$) [56].

Using BLS, we studied the influence of $t_{Co}$ and $t_W$ on iDMI. Two series of samples were chosen: (*i*) with the fixed $t_W$ and different nominal thickness of the ferromagnetic layer - Ru(10)/Co($t_{Co}$)/W(0.23)/Ru(2); (*ii*) with the fixed $t_{Co}$ and various $t_W$ - Ru(10)/Co(1)/W($t_W$)/Ru(2). Figure 5(b) shows the dependence of $D_{eff}$ on the $t_{Co}$. For the thickness range of $t_{Co}$ below 1.0 nm, where PMA exists, DMI has non-linear dependence. At $t_{Co}$>1.17 nm the in-plane magnetic anisotropy appears and the DMI value significantly decreases down to -0.4 erg/cm$^2$ for $t_{Co}$=1.5 nm. The interfacial origin of iDMI is proven by the linear dependence of $D_s$=$f(1/(t_{Co}-A))$ for $t_{Co}$ > 1nm (see the inset in Fig.5(d)). The peak iDMI value ($D_{eff}$=-2.1 erg/cm$^2$) was observed for $t_{Co}$=1 nm. For this particular Co thickness we studied the dependences of $D_{eff}$ and $D_s$ as a function of $t_W$, Fig.5(c). For the quasi-symmetrical Ru(10)/Co(1)/Ru(2) the value of $D_{eff}$ is -0.27 erg/cm$^2$. The existence of the non-zero negative iDMI points out the different morphological quality (roughness, intermixing depth) of the Ru/Co and Co/Ru interfaces, as it was revealed for Pt/Co/Pt trilayers [22, 23]. We found for Ru/Ru films that if the thickness of the Ru capping decreases down to 0.5 nm, the value of $D_{eff}$ increases up to -1.31 erg/cm$^2$, possibly due to the partial oxidation of the Co layer through the very thin and structurally not continuous Ru capping layer [57]. This fact allows us to assume, that in this particular case the iDMI contribution from the top Co/Ru interface is negligible ($D_{Co/Ru}\approx$ 0) and $D_{eff}$ =-1.31 erg/cm$^2$ is mostly induced by the bottom Ru/Co interface. It means, that in case of the quasi-symmetric Ru/Co/Ru films with the 2-nm thick Ru capping, the top Co/Ru interface has the positive DMI with the value defined as $D_{Co/Ru}$=$D_{eff(Ru(10)/Co(1)/Ru(2))}$ - $D_{Ru/Co}$ =-0.27+1.31=1.04 erg/cm$^2$. A scheme, illustrating the partial cancelation of DMI in the quasi-symmetrical Ru/Co/Ru film, is represented in Fig.5(d).

The introduction of the W interlayer with thickness up to 0.24 nm allows to considerably enhance the structural inversion asymmetry in Ru/Co/W/Ru films resulting in more than 10 times increase of $D_{eff}$. The peak value -3.1 erg/cm$^2$ was observed for $t_W$=0.24 nm. In the range 0.24 nm<$t_W$<0.31 nm we found the abrupt fall of $D_{eff}$ down to -0.74 erg/cm$^2$ accompanied by the decrease of $K_{eff}$ as shown on Fig.3(d).

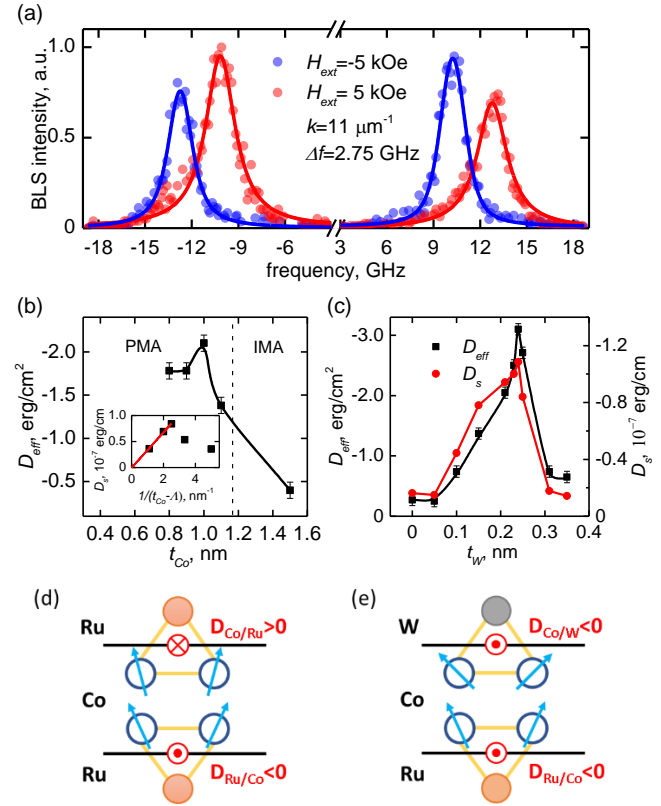

FIG. 5. (a) BLS spectra recorded for the Ru(10)/Co(1)/W(0.25)/Ru(2) film. Symbols refer to the experimental data and solid lines are the Lorentzian fits. (b) Dependence of $D_{eff}$ on the ferromagnetic layer thickness ($t_{Co}$) at the fixed $t_W$=0.23 nm. The inset demonstrate the linear dependence $D_s$=$f(1/(t_{Co}-A))$ for the relatively thick films. (c) $D_{eff}$ and $D_s$ values as a function of $t_W$ at the fixed $t_{Co}$=1 nm. (d)

We suppose that the sharp increase of $D_{eff}$ is provoked by the additive effect of Ru/Co and Co/W interfaces, due to Ru/Co and W/Co interfaces have opposite signs of DMI: negative for Ru and positive for W. It means that $D_{Ru/Co}$<0 and $D_{Co/W}$<0, as illustrated in Fig.5(e). We determined the iDMI value of the Co/W interface for Ru(10)/Co(1)/W(0.24)/Ru(2) as $D_{Co/W}$=$D_{eff}$-$D_{Ru/Co}$=-3.1+1.31=-1.79 erg/cm$^2$. The decrease of $D_{eff}$ for $t_W$ > 0.24 nm is due to the increase of the magnetically dead layer on the Co/W interface (Fig.2(b)), which leads to the consequent degradation of the Co layer.

In order to compare the DMI values for different systems the interfacial constant $D_s$ has to be calculated for the effective thickness of the ferromagnetic layer taking into account the accurately defined magnetically dead layers of the bottom and top interfaces, Fig.5(c). Otherwise, the $D_s$ value can be significantly overestimated, when the nominal thickness is in use. It is important, that Ru/Co/W/Ru films with 0.8 nm<$t_{Co}$<1.17 nm and 0.15 nm<$t_W$< 0.3 nm can be used for skyrmion stabilization [32], due to they have the iDMI values, which are larger than the critical energy density for the formation of the chiral domain walls written as $D_{cr} = \frac{4}{\pi}\sqrt{AK_{eff}}$ [58], where $A$ is the exchange stiffness constant. For $A$=1.6×10⁻⁶ erg/cm and $K_{eff}$=0.6×10⁶ erg/cm³, one can find that $D_{cr}$=1.25 erg/cm³. Accordingly to [32], our modified Ru/Co/W/Ru films can be potentially used not only for the homochiral Neel domain walls and isolated skyrmions nucleation, but mainly for the skyrmion lattice stabilization, making the realization of the skyrmionic memory and domain wall based devices closer.

## VII. OBSERVATION OF SKYRMIONS IN TRILAYER FILMS

For the creation and stabilization of skyrmions in thin films a unique combination of the magnetic parameters, such as PMA, Heisenberg exchange, $M_s$ and iDMI, is required [55, 59-61]. Due to the lack of the iDMI strength, a multilayer configuration of films with the magnetostatically coupled layers is usually used for the skyrmion stabilization and manipulation [3]. We found by means of polar Kerr effect microscopy that iDMI stabilized skyrmions in the Ru/Co/W/Ru films with $t_{co}$=1.1 nm, which have relatively weak PMA ($K_{eff}$≈0.3×10⁶ erg/cm³). Figure 6 shows the out-of-plane field driven evolution of the magnetic domain structure in the Ru(10)/Co(1.1)/W(0.25)/Ru(2) film. The initial saturation state was achieved in the positive field +$H_s$ and it is not presented in Fig.6. Due to the limited resolution of Kerr microscopy, skyrmions with the size less than 1 μm became visible in the field $H$ = -35.2 Oe (skyrmions are pointed out by red arrows in Fig.6(a)). With the increasing field up to $H$=-36 Oe, besides the nucleation of new skyrmions the fast growing magnetic domains appeared, Fig.6(b). At the same time a part of skyrmions grew up into the non-circular domains with uneven edges (marked by yellow arrows in Fig.6(b)). In the filed $H$>-37 Oe, the size and number of domains increased, Fig.6(c-f).

To prove the skyrmion formation in the Ru/Co/W/Ru trilayers, we performed the micromagnetic simulations of the 4 × 4 μm² area of films using the continuous boundary conditions. Recently, we have demonstrated that at the Ru buffer layer thickness of 10 nm the Co films are polycrystalline [39]. It leads to the local variation of magnetic properties and, finally, can cause in the increased critical fields, namely, the coercive force due to the domain wall pinning. Therefore, magnetic inhomogeneities in disordered films can pin skyrmions or change their shape and size [62-64]. In the simulations we considered the variation of the crystallographic anisotropy $K_l$= 0.3 × 10⁶ erg/cm³ on the grains with size of 20 nm [39, 63]. The orientation of the easy axis of magnetization was assigned randomly for each crystallite and the value of $K_l$ was varied within ±10%. We used the following parameters: /$D_{eff}$/=1.0 erg/cm² (which is larger than $D_{cr}$=0.88 erg/cm² for this film), $M_s$=1300 emu/cm³, $K_{eff}$=0.3×10⁶ erg/cm³, $A$=1.6 ×10⁻⁶ erg/cm, and the cell size 4×4×5 nm³.

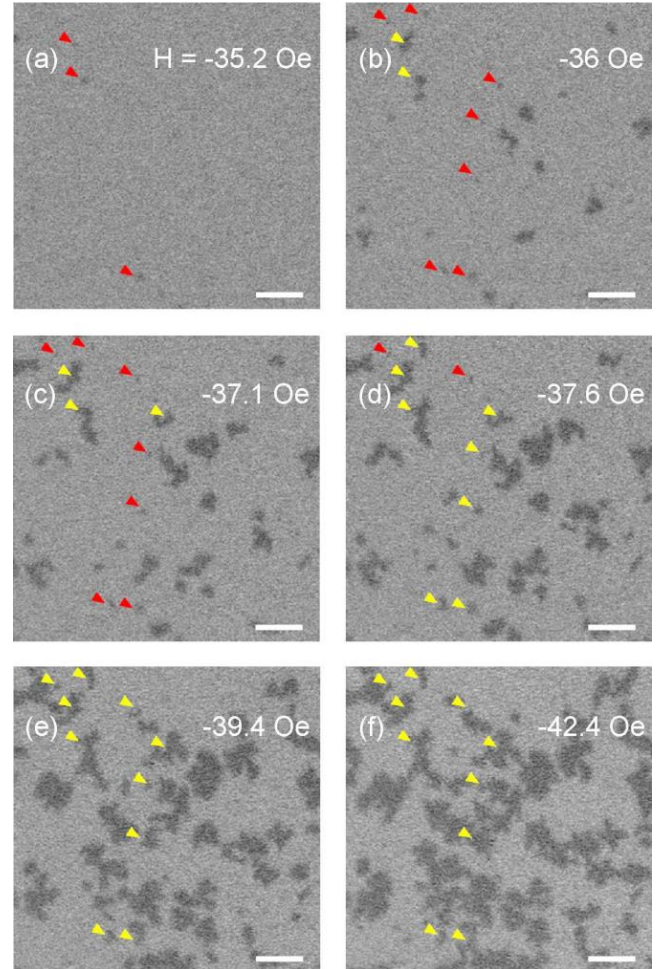

FIG. 6. Magnetic domain structure of the Ru(10)/Co(1.1)/W(0.25)/Ru(2) film captured in the out-of-plane field by Kerr microscopy. The red arrows indicate skyrmions, which grow and form domains marked by the yellow arrows. The scale bar is 5μm.

At the beginning, the six stable skyrmions were created in zero field, Fig.7(a). One can see that the shape and size of skyrmions are different because of the magnetic inhomogeneities. The average skyrmion diameter is 224 nm. All the skyrmions have the skyrmion number 1. With the rising magnetic field up to 30 Oe, the shape of 4th and 5th skyrmions significantly changed, meantime the size of 2nd and 3rd skyrmions increased. In the fields of 40 and 50 Oe only 3rd skyrmion became unchanged. Further increase of the field (Fig.7(e,f)) promoted the sharp growth of skyrmions and their transformation into large domains. The very similar behavior was observed experimentally, Fig.6. We found that without consideration of disorders in films, skyrmions were unstable in zero magnetic field and they immediately transformed into domains.

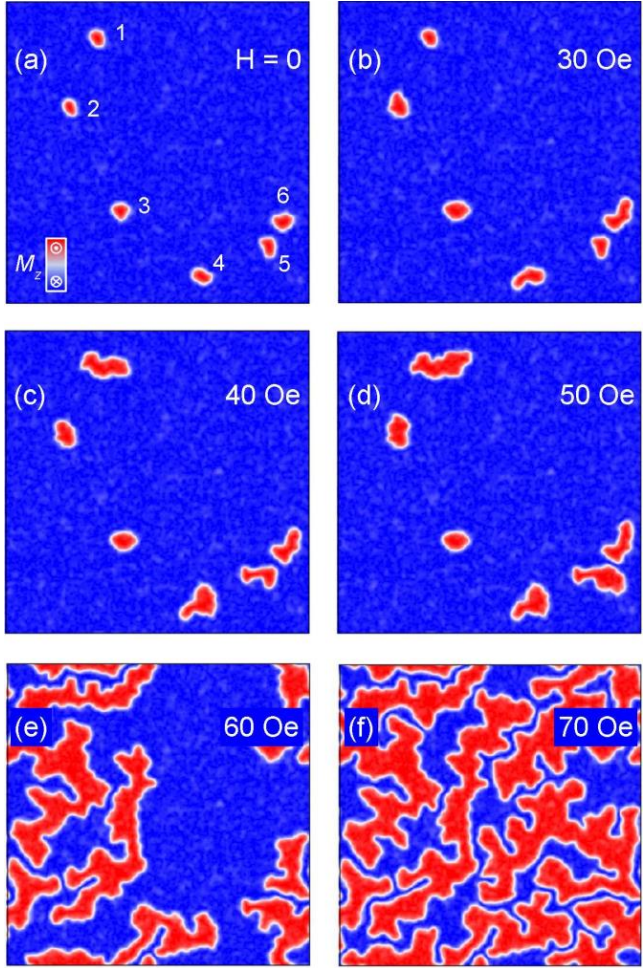

**FIG.7.** Simulated images of the out-of-plane field driven domain structure of the Ru(10)/Co(1.1)/W(0.25)/Ru(2) film. The image size is $4 \times 4 \ \mu m^2$.

## VIII. CONCLUSIONS

Our study of the broken structural inversion symmetry Ru/Co/W/Ru films has unveiled the thicknesses of Co and W layers for which the strong PMA exists and the iDMI remarkably enhances (up to -3.1 erg/cm$^2$) thanks to the additive chiral interaction if to compare with the quasi-symmetrical Ru/Co/Ru trilayers. The introduction of the W interlayer into the top Co/Ru interface decreases the elastic strains in the ferromagnetic layer, smooths the Co surface and dramatically decreases the coercive force. The created unique combination of structural and magnetic parameters enables to nucleate and stabilized skyrmions with the diameter of 200 nm in the films with only one ultrathin ferromagnetic layer. Our findings show the high potential of ruthenium as an alternative to platinum and open up the way for the extensive interface engineering at the nanoscale in order to tune magnetic properties and to strengthen spin-orbit effects for the future spin-orbitronic applications, especially, for skyrmionic devices.

## ACKNOWLEDGMENTS

This work has been partially supported by the by the Russian Foundation for Basic Research (grant 17-52-45135), by the Russian Ministry of Education and Science under the state task (3.5178.2017/8.9 and 3.4956.2017), by the Grant program of the Russian President (MK-2643.2017.2) and by the Brain Pool Program (172S-2-3-1928) through the Korean Federation of Science and Technology Societies (KOFST) funded by the Ministry of Science, ICT and Future Planning.